%% file: main.tex
\PassOptionsToPackage{unicode,pdfusetitle,bookmarksnumbered}{hyperref}
\RequirePackage{hyperref}
\documentclass[sigconf,nonacm]{acmart}
\usepackage{etex}
\usepackage{subcaption}
\usepackage{booktabs}
\usepackage[noend]{algpseudocode}
\usepackage{adjustbox}
\usepackage{standalone}
\usepackage{pgf-pie}
\usepackage{comment}
\usepackage{threeparttable}

\let\terms\undefined

\acmBooktitle{}

\acmConference[]{}{}{}

\input{preamble}
\title{Content Hidden Behind Execution: Analyzing Public Scratch Projects at Runtime}

\author{Yuan Si}
\affiliation{%
  \institution{University of Waterloo}
  \city{Waterloo}
  \country{Canada}}

\author{Yufeng Lin}
\affiliation{%
  \institution{Pui Ching Middle School Macau}
\city{Macau}
\country{China}}

\author{Daming Li}
\affiliation{%
  \institution{Independent Researcher}
  \city{San Mateo}
  \country{USA}}

\author{Hanyuan Shi}
\affiliation{%
  \institution{Independent Researcher}
  \city{Hangzhou}
  \country{China}}

\author{Yang Shi}
\affiliation{%
  \institution{Utah State University}
  \city{Logan}
  \country{USA}}

\author{Jialu Zhang}
\authornote{Corresponding author.}
\affiliation{%
  \institution{University of Waterloo}
  \city{Waterloo}
  \country{Canada}}

\begin{document}

\begin{abstract}
\input{sec/abs}
\end{abstract}

\keywords{Scratch, computing education, block-based programming, project curation, runtime behavior}
\maketitle

\input{sec/intro}
\input{sec/BGMotivation}
\input{sec/related}
\input{sec/taxonomy}
\input{sec/study}
\input{sec/evaluation}
\input{sec/discussion}
\input{sec/conclusion}

\bibliography{scratch}

\end{document}

%% file: preamble.tex
\usepackage{graphicx}
\usepackage{amsmath}

\usepackage{xspace}
\usepackage{footnote}
\usepackage{wrapfig}  

\usepackage{amsfonts}
\usepackage{natbib}
\usepackage{hhline}
\usepackage{multirow}
\usepackage{setspace} 
\usepackage{url}
\usepackage{booktabs}
\usepackage{xcolor}
\usepackage{lipsum}
\usepackage{courier}
\usepackage{listings}
\usepackage{caption}
\usepackage{colortbl}
\usepackage{arydshln} 
\usepackage{makecell} 
\usepackage{tikz}
\usepackage{xcolor}
\usetikzlibrary{calc}
\usepackage{varwidth}
\usepackage{natbib}
\usepackage{algorithm}
\usepackage[noend]{algpseudocode}
\usepackage{subcaption}
\usepackage{wrapfig}

\usepackage[most]{tcolorbox}
\usepackage{tabularx}

\newcolumntype{Y}{>{\raggedright\arraybackslash}X}

\usetikzlibrary{fit,arrows,calc,positioning,quotes}
\usetikzlibrary{decorations.pathreplacing}

\usepackage{mathtools}
\usepackage{fancyvrb}
\usepackage{textcomp}
\usepackage{stmaryrd}

\tikzset{
  -|-/.style={
    to path={
      (\tikztostart) -| ($(\tikztostart)!#1!(\tikztotarget)$) |- (\tikztotarget)
      \tikztonodes
    }
  },
  -|-/.default=0.5,
  |-|/.style={
    to path={
      (\tikztostart) |- ($(\tikztostart)!#1!(\tikztotarget)$) -| (\tikztotarget)
      \tikztonodes
    }
  },
  |-|/.default=0.5
}

\usetikzlibrary{shapes.geometric}

\long\def\com#1{}

\newcommand{\para}[1]{\noindent {\bf #1}}

\newcommand{\squishlist}{
   \begin{list}{$\bullet$}
    { \setlength{\itemsep}{0pt}      \setlength{\parsep}{3pt}
      \setlength{\topsep}{3pt}       \setlength{\partopsep}{0pt}
      \setlength{\leftmargin}{3.5mm} \setlength{\labelwidth}{1em}
      \setlength{\labelsep}{0.5em} } 
}

\newcommand{\squishend}{
    \end{list}  }

\def\BibTeX{{\rm B\kern-.05em{\sc i\kern-.025em b}\kern-.08em
    T\kern-.1667em\lower.7ex\hbox{E}\kern-.125emX}}


%% file: sec/abs.tex
Public Scratch projects are reused in computing education as classroom examples, remix sources, open-exploration materials, and research data. Curation often begins with titles, thumbnails, descriptions, tags, and remix links, but Scratch projects are executable learning artifacts. Content affecting age appropriateness can appear only after execution, gameplay progression, a failure state, user interaction, costume switching, audio playback, or a hidden event trigger.

We study \emph{runtime-revealed sensitive content} as a computing education curation challenge: educators and researchers need runtime evidence about what students may encounter when Scratch projects are used in these settings. We introduce a runtime-aware annotation scheme that separates content category, risk level, evidence channel, reveal mechanism, and annotation confidence. Using this scheme, we conducted an audit of 500 public Scratch projects sampled from curated candidates, taxonomy-guided keyword search, and follow-up exploration of project clusters surfaced during review.

In this audit, 467 of 500 projects (93\%) required runtime exploration beyond static metadata to surface the safety-relevant signal; 387 (77\%) required interaction, gameplay progression, failure states, or hidden-asset and code inspection. As a targeted classroom and research curation audit, the study characterizes reveal mechanisms in a selected corpus rather than estimating platform-wide prevalence or making platform-level safety claims. The results show metadata-only screening leaves key evidence unresolved in executable youth media. By separating content type, severity, evidence location, and reveal pathway, this work supports classroom project selection, student exploration practices, dataset construction, and educator-facing screening tools for block-based programming communities.

%% file: sec/intro.tex
\section{Introduction}
\label{sec:intro}

Scratch supports computing education through projects that learners create, share, inspect, and remix. In constructionist and creative-learning traditions, learners build personally meaningful artifacts and learn through peers, feedback, and reflection~\cite{papert1980mindstorms,resnick2009scratch,resnick2017lifelong}. Project-based learning similarly treats meaningful artifacts as a basis for motivation and understanding when teachers, peers, and tools provide usable scaffolding~\cite{blumenfeld1991motivating,barron1998doing}. Scratch makes these ideas concrete in a public online community where young people publish interactive media and learn from projects made by others~\cite{scratch_platform,hill2017scratchdataset,dasgupta2016remixing,cheng2022interest}.

Public sharing makes project selection a student-facing educational decision. Teachers may choose public projects as worked examples, ask students to browse for inspiration, or use remixing in project-based activities, but students are often the eventual users: they browse, play, inspect, and remix projects during learning. Researchers may also collect public projects for datasets about novice programming, debugging, creativity, assessment, or computational thinking. In each pathway, a Scratch project is a learner-facing material, not only a code artifact.

We use \emph{curation} to mean selecting and documenting public projects before learners encounter them as classroom examples, open-exploration materials, remix sources, or dataset entries. Curation often begins with the project page. Titles, thumbnails, instructions, descriptions, tags, social signals, and remix links help educators and dataset builders decide which resources deserve attention. Work on digital learning-object repositories and teacher resource exchanges shows that metadata and evaluative cues shape discovery, judgment, and reuse~\cite{currier2004metadata,abramovich2013metadata}. Scratch pages provide similar cues, but the executed project can contain evidence that the page never shows. The central screening question is whether the curator has enough evidence about the learner's experience, not whether the page surface contains an obvious warning sign.

We call this phenomenon \emph{runtime-revealed sensitive content}: sensitive or age-inappropriate content that appears only after execution, gameplay progression, a failure state, a timed delay, user input, costume switching, audio playback, or a hidden event trigger. A project page can look benign while normal play reveals a jump scare, an intense sound, a threatening scene, or hidden content. Some learners may also deliberately seek scary or edgy projects and share them with peers, so curation should support guidance rather than assume exposure is always accidental. A broad theme label such as horror or combat is also insufficient because classroom appropriateness depends on severity, presentation, evidence location, reveal pathway, and likely learner encounter.

This distinction matters first for students' learning experience, and then for the educators and researchers who structure that experience. Scratch community guidelines, reporting, moderation, and digital-citizenship resources provide a platform-level safety layer~\cite{scratch_guidelines,scratch_digital_citizenship}. Educators still need to judge whether a project fits a particular age group, assignment, and classroom norm. Dataset builders also need to report what screening examined and what evidence supports labels or exclusions. Metadata-only review leaves those decisions under-specified when the relevant signal appears during execution.

This paper presents a runtime-aware audit of 500 public Scratch projects. The audit characterizes how safety-relevant content is encoded, how it becomes visible or audible, and what evidence is needed to annotate it for classroom reuse, student exploration, remix contexts, and dataset screening. The sample combines curated candidates, taxonomy-guided keyword search, and follow-up exploration of project clusters surfaced during review. For each project, annotators inspected metadata, executed the project, performed standard interactions, and recorded whether the primary safety-relevant signal was visible statically or required runtime exploration, interaction, gameplay progression, failure states, or code/asset inspection.

We organize the study around three research questions:
\begin{description}
    \item[RQ1:] What forms of safety-relevant content appear in targeted public Scratch project pathways used to stress-test classroom, student-exploration, remix, and dataset curation?
    \item[RQ2:] Which safety-relevant signals are visible from project metadata, and which require execution, interaction, gameplay progression, failure states, or code/asset inspection?
    \item[RQ3:] What category boundaries, evidence channels, and reveal mechanisms make Scratch project curation difficult for educators and researchers?
\end{description}

The remainder of the paper motivates the problem with concrete examples, situates it in related work on Scratch, educational resource selection, youth safety, and interactive media classification, then presents the annotation scheme, audit method, findings, and implications for runtime-aware classroom and dataset curation.

%% file: sec/BGMotivation.tex
\section{Background and Motivating Examples}
\label{sec:background}

Scratch projects sit at the intersection of programming and media. A single project can combine code, sprites, costumes, sounds, text, animation, and user interaction. It can also be shared, remixed, collected in studios, and discovered by learners browsing a public community. In this paper, content safety means whether that executable experience is appropriate for a particular learning activity, age group, and classroom norm; it is not a claim that Scratch itself is unsafe. We therefore treat Scratch project review as a runtime-aware educational curation problem where the relevant artifact is an executable experience that learners may play, inspect, remix, and encounter through peers.

We group several motivating examples below to make this issue concrete for computing-education readers. The examples illustrate why unexpected runtime content can matter in a learning setting, while remaining separate from any claim about platform-wide prevalence. They also show why content-safety analysis cannot stop at titles, thumbnails, or descriptions: Scratch projects unfold through progression, timing, interaction, failure states, remix context, audio playback, and hidden code paths.

\paragraph{Benign-framed maze and jump-scare projects.}
For example, our motivating set included a maze game and a cheese-collection game whose page surfaces did not clearly announce the later scare. In the maze game, the learner guides a small object through simple levels, and the opening state gives little reason to expect an intense scare sequence. After gameplay progression, a runtime event switches away from the maze, shows a sudden visual change, plays a scream, and resumes the game. In the cheese-collection game, ordinary play can lead to a failure state that changes the backdrop to a startling face image, flashes the screen, hides the sprites briefly, and plays a loud scream. In both projects, the learner reaches the reveal by doing what the project asks them to do. The risky moment is produced by normal play rather than announced by the page surface.

\paragraph{FNAF-inspired horror projects.}
Scratch also contains Five Nights at Freddy's style fan games that appear through studios, remix chains, and genre collections; ESRB describes the commercial series as puzzle/horror games with frequent screams and jump scares~\cite{esrb_fnaf_core}. These clusters show that curation evidence can extend beyond a single project page. A mild fan game, a tense survival game, and a jump-scare-heavy remake can share a franchise label while creating different classroom concerns. Risk depends on intensity, presentation, audio, progression, and runtime behavior.

\paragraph{War and tower-defense projects.}
Public remix pages also include strategy or tower-defense games with stylized combat mechanics, as well as war-themed projects whose titles reference military conflict, tanks, or defense. Tower-defense mechanics are not themselves the concern. These projects motivate the distinction between content category and risk level: stylized or fantasy conflict can be low risk, while explicit threats, graphic injury, hate framing, or realistic gore can raise the risk level~\cite{esrb2024ratings,pegi2024ratings}. Treating all combat-themed projects as equally unsafe overstates risk and obscures differences that matter for classroom curation.

\paragraph{School-horror and Baldi-style remix clusters.}
Scratch includes remix clusters around Baldi-style games, soundboards, and school-horror projects; the official Baldi's Basics app description frames the game as meta horror set in a school-like environment~\cite{baldi_googleplay}. These cases combine school settings, horror framing, chase mechanics, hostile characters, sounds, and bullying-like language. The annotation challenge is to distinguish fictional school-horror mechanics from harassment or targeted bullying, especially when the evidence appears only through play, interaction, or audio.

Together, these cases motivate three annotation choices. Categories need to be multi-label because projects can combine horror, violence, profanity, harassment, and other themes. Risk levels need to remain separate from categories because themes such as combat or horror can be appropriate in some contexts. Annotations also need to record evidence channel and reveal mechanism because a teacher or dataset builder needs to know not only what content exists, but how a learner would encounter it: in metadata, runtime behavior, interaction, remix context, audio, or hidden code paths.

%% file: sec/related.tex
\section{Related Work}
\label{sec:related}

\para{Scratch as a public learning community.} Scratch has been studied as both a block-based programming environment and a public learning community. Constructionist and creative-learning work emphasizes meaningful artifacts, sharing, remixing, and reflection~\cite{papert1980mindstorms,resnick2017lifelong}. Community-level studies examine public activity, remix pathways, and informal learning from peer-produced projects~\cite{hill2017scratchdataset,dasgupta2016remixing,cheng2022interest,si2025stitch,si2026ecoscratch}. This literature makes the exposure scenario in our study educationally plausible: learners encounter public Scratch projects through browsing, inspiration, reuse, remix, and classroom activity. Public projects operate as examples, play materials, code resources, and social artifacts at the same time. A curation framework for Scratch must account for all of these roles because a learner may experience the project before, during, or after reading its code.

\para{Student-facing resource selection and metadata-mediated curation.} Formal resource selection often happens through teachers, who rely on repositories, previews, descriptions, ratings, and metadata~\cite{AnastasiiaKoli,artser2026debugging,zhang2025tle,gu2026contextaware}. Metadata quality shapes discovery and reuse in learning-object repositories~\cite{currier2004metadata}, and evaluative metadata influences which resources teachers download and use~\cite{abramovich2013metadata}. Students, however, may also encounter resources less systematically through browsing, studios, remix links, or peer sharing. Scratch pages provide comparable surface cues, useful for teacher triage and often the only cues a learner sees before pressing the green flag. The curation task changes when the resource is an executable program: a preview is incomplete unless it records what was actually run, heard, clicked, failed, or inspected.

\para{Scratch analysis, feedback, and runtime support.} Computing-education research has developed tools for analyzing Scratch programs~\cite{li2026raven,lin2026scratchworld,si2026checkedprogramrecoveryexecution,si2026scratchlenslensparametricbehavioralequivalence}. Static and structural systems inspect projects for computational-thinking constructs, misconceptions, bugs, and code smells~\cite{brennanresnick2012ct,fraser2021litterbox}. Runtime-oriented systems use execution and tests to support debugging and functional feedback~\cite{deiner2024nuzzebug,feldmeier2024blocktesting,si2025viscratch,si2026certificatecarryingtransformationeventdrivenblock,si2026schedcheckschedulerobustnessanalysiseventdriven}. These systems show that Scratch code and runtime behavior can be instrumented for learning research. They also clarify the boundary of our contribution: the problem is not whether Scratch can be executed or analyzed, but which runtime evidence is needed when the target judgment concerns age appropriateness and educational reuse. Our work uses runtime evidence for that curation task: documenting whether public executable projects are appropriate for classroom reuse, open exploration, remix support, or dataset construction.

\para{Youth online safety and platform moderation.} Youth online safety research has examined harassment, bullying, hate speech, privacy leakage, self-harm exposure, inappropriate content, and platform governance~\cite{livingstone2014harms,gillespie2018custodians}. Work on youth-facing game and user-generated-content platforms shows that risks can appear through play, social interaction, promotional imagery, and moderation-evasive behavior~\cite{zhang2025dangerous}. Scratch addresses safety through community guidelines, moderation, reporting, and digital-citizenship resources~\cite{scratch_guidelines,scratch_digital_citizenship}. Educational curation adds a classroom and research layer: a project can be unsuitable for a specific age group, assignment, or open-exploration activity even when similar themes remain appropriate elsewhere.

\para{Interactive media classification and runtime evidence.} Game and interactive-media research treats user experience as emerging from rules, mechanics, timing, progression, and audiovisual presentation~\cite{hunicke2004mda}. Age-rating systems such as ESRB and PEGI distinguish age categories, content descriptors, interactive elements, and contextual severity~\cite{esrb2024ratings,esrb2026process,pegi2024ratings}. ESRB's process for physical games asks publishers to disclose relevant content and provide video of typical gameplay and extreme content~\cite{esrb2026process}. This supports our central premise: interactive media require evidence about what a user can encounter during play. Scratch differs because projects are youth-generated, continuously edited, remixed, and discovered through a public learning community.

\para{Runtime-aware curation for computing education.} Prior work establishes Scratch as a public learning community~\cite{resnick2009scratch}, studies metadata for educational selection, instruments Scratch execution~\cite{si2026scratcheval}, and examines exposure context in interactive media. Missing is a framework for Scratch projects as executable learning materials whose appropriateness depends on runtime pathways invisible to metadata-only review. Safety-design work favors agency-preserving, collaborative guidance for minors~\cite{wisniewski2017middleground}. This gap motivates our runtime-aware annotation scheme and audit of 500 public Scratch projects.

%% file: sec/taxonomy.tex
\section{Taxonomy and Annotation Scheme}
\label{sec:taxonomy}

We developed a runtime-aware annotation scheme for public Scratch projects. The scheme describes five aspects: what safety-relevant content appears, how severe it is, where the evidence is observed, how the content is revealed, and how strongly the label is supported. It is designed for educational curation and research documentation; moderation decisions remain a platform-governance matter.

The scheme was developed in two passes. First, we translated content axes from youth online-safety work, Scratch community guidance, and interactive-media rating systems into candidate category and severity labels~\cite{livingstone2021fourcs,scratch_guidelines,esrb2024ratings,pegi2024ratings}. Second, during pilot mapping of curated Scratch projects, we refined the operational dimensions whenever the relevant cue depended on running the project, hearing audio, triggering an event, or inspecting assets and code. This process led us to separate content category, risk level, evidence channel, reveal mechanism, and confidence instead of using a single appropriateness label.

\subsection{Content Categories}

Content categories are multi-label because Scratch projects often combine horror, violence, harassment, social or political content, and moderation-conflict signals. Table~\ref{tab:categories} summarizes the nine categories used in this study. Categories C1--C8 follow common youth-safety axes; C9 captures Scratch-specific signals in which a project discourages reporting, challenges moderation, or reposts contested material. Thumbnail/runtime mismatch is captured separately as reveal mechanism M7 (Table~\ref{tab:idcatalog}).

\begin{table}[t]
\centering
\caption{Content categories for safety-relevant Scratch content.}
\label{tab:categories}
\footnotesize
\setlength{\tabcolsep}{3pt}
\renewcommand{\arraystretch}{1.02}
\begin{tabularx}{\linewidth}{@{}l p{0.24\linewidth} X@{}}
\toprule
ID & Category & Definition \\
\midrule
C1 & Violence / Threat & Physical harm, weapons, combat, pursuit, or violent threats. \\
C2 & Horror / Survival & Fear, jump scares, hostile pursuit, hiding, or survival pressure. \\
C3 & Gore / Blood & Blood, injury, body damage, or graphic violence. \\
C4 & Sexual / NSFW & Nudity, sexual imagery, explicit references, or adult framing. \\
C5 & Hate / Discrimination & Identity-targeted hate, exclusion, or dehumanization. \\
C6 & Harassment / Bullying & Targeted insults, humiliation, coercion, or threats. \\
C7 & Self-harm / Suicide & Self-harm, suicide, severe despair, or encouragement of self-harm. \\
C8 & Political / Societal & Political figures, elections, war, conflict, or sensitive public issues. \\
C9 & Moderation Conflict / Report Avoidance & Reporting discouragement, moderation challenges, or reposting contested material. \\
\bottomrule
\end{tabularx}
\end{table}

Category labels describe content type; risk levels describe severity. A stylized tower-defense game and a graphic violent scene both involve violence with different risks. Political or societal content is labeled for contextual sensitivity, not high risk by default.

\subsection{Risk, Evidence, Reveal, and Confidence}

Beyond category labels, each annotation records four operational dimensions, summarized in Table~\ref{tab:operational}. The two multi-label dimensions, evidence channel and reveal mechanism, use identifiers (E0--E6 and M0--M8) catalogued in Table~\ref{tab:idcatalog}. These dimensions separate content type from severity and record whether evidence comes from metadata, assets, runtime behavior, interaction, public context, or code-level inspection.

\begin{table}[t]
\centering
\caption{Operational annotation dimensions.}
\label{tab:operational}
\footnotesize
\setlength{\tabcolsep}{3pt}
\renewcommand{\arraystretch}{1.02}
\begin{tabularx}{\linewidth}{@{}p{0.29\linewidth} X@{}}
\toprule
Dimension & Labels / scale \\
\midrule
Risk level & Project-level severity 0--4: normal; sensitive but acceptable; mildly age-inappropriate; clearly inappropriate; high-risk or likely policy-violating. \\
Evidence channel & E0--E6: where the signal was observed; see Table~\ref{tab:idcatalog}. \\
Reveal mechanism & M0--M8: how the signal becomes visible or audible; see Table~\ref{tab:idcatalog}. \\
Annotation confidence & Low, Medium, High, Confirmed. Confirmed requires direct runtime, source, or asset evidence; lower levels indicate weaker or ambiguous evidence. \\
\bottomrule
\end{tabularx}
\end{table}

\begin{table}[t]
\centering
\caption{Identifier catalogue for evidence channels and reveal mechanisms.}
\label{tab:idcatalog}
\footnotesize
\setlength{\tabcolsep}{3pt}
\renewcommand{\arraystretch}{1.0}
\begin{tabularx}{\linewidth}{@{}l X@{}}
\toprule
ID & Description \\
\midrule
\multicolumn{2}{@{}l}{\emph{Evidence channels}} \\
E0 & Metadata: title, instructions, description, tags, remix metadata \\
E1 & Thumbnail or rendered first frame \\
E2 & Static assets: costumes, sprites, backdrops, sounds, embedded text \\
E3 & Source-code structure: blocks, broadcasts, clones, hats, timing \\
E4 & Passive runtime trace after green-flag start \\
E5 & Interactive runtime trace after click, key, mouse, or typed input \\
E6 & Public contextual evidence: forum, studio, profile, remix chain \\
\midrule
\multicolumn{2}{@{}l}{\emph{Reveal mechanisms}} \\
M0 & Static / metadata-level \\
M1 & Passive runtime \\
M2 & Interaction-triggered \\
M3 & Timed or delayed \\
M4 & Broadcast / state-triggered \\
M5 & Composite or hidden-asset reveal \\
M6 & Audio-dominant \\
M7 & Thumbnail / runtime mismatch \\
M8 & Source-only or asset-only latent content \\
\bottomrule
\end{tabularx}
\end{table}

\subsection{Boundary Rules}
\label{subsec:boundaries}

Boundary rules reduce over-labeling. Categories are assigned from observable evidence rather than inferred creator intent. Stylized conflict is distinguished from graphic violence. Harassment requires targeting an identifiable user, creator, or group. Political or societal themes are labeled for contextual sensitivity but do not imply high risk. When evidence is ambiguous, annotators choose the lowest risk level supported by the evidence and record uncertainty.

%% file: sec/study.tex
\section{Study Design}
\label{sec:study}

We conducted a seed-based audit of 500 public Scratch projects. The goal was to test whether runtime-revealed sensitive content can be categorized and annotated at scale, not to estimate platform-wide prevalence. We intentionally sampled projects likely to contain sensitive, borderline, or moderation-relevant content so the annotation scheme would encounter difficult cases across the taxonomy.

\subsection{Seed Set}

Candidate projects came from three sources (Table~\ref{tab:seed}). First, a curated pool of 50 projects from prior taxonomy mapping supplied examples spanning C1--C9. Second, targeted keyword search against public Scratch search pathways supplied candidates from terms derived from the nine-category taxonomy, such as ``shooter,'' ``stick war,'' ``FNAF,'' ``Granny,'' ``scary maze,'' ``stop bullying,'' ``stand with Ukraine,'' and ``don't report''~\cite{scratch_platform}. Third, follow-up exploration sampled two high-density public project clusters surfaced during early review, with repeated C9 moderation-conflict and C6-like harassment signals. Across sources we screened a larger candidate pool ($N \approx 1{,}458$ projects fetched and pre-classified) and retained 500 final entries after runtime processing, franchise-clone deduplication, and visual review.

\begin{table}[t]
\centering
\caption{Seed set composition for the 500-project audit.}
\label{tab:seed}
\footnotesize
\setlength{\tabcolsep}{3pt}
\renewcommand{\arraystretch}{1.02}
\begin{tabularx}{\linewidth}{@{}X r@{}}
\toprule
Source & Count \\
\midrule
Curated input candidates from prior taxonomy mapping & 50 \\
Targeted keyword search guided by C1--C9 taxonomy & 395 \\
Project-cluster follow-up from two high-density public clusters & 55 \\
\midrule
Total & 500 \\
\bottomrule
\end{tabularx}
\end{table}

\subsection{Annotation Procedure}
\label{subsec:annotation}

We used an LLM-assisted first-pass annotation pipeline followed by human-supervised checkpoint review and final adjudication. For each project, a headless \texttt{scratch-vm} harness executed the project with a green flag, a passive runtime window, and a synthesized interaction timeline that fired declared hat triggers plus a fixed navigation set~\cite{scratchvm}. A multimodal LLM then reviewed three rendered keyframes (start, middle, end), project metadata, and a structured trace summary. The reviewer produced a draft annotation along the five dimensions in Section~\ref{sec:taxonomy}, applied the boundary rules in Section~\ref{subsec:boundaries}, and flagged uncertain or high-impact cases for human review. When frames were uninformative, the reviewer fell back to metadata-only classification and lowered confidence.

A lead researcher supervised the audit through periodic checkpoint reviews. The supervisor was the project lead who created the codebook and completed the prior taxonomy-mapping pass used to form the curated seed pool; this role provided continuity for adjudication but is not a substitute for independent reliability measurement. At each checkpoint, the supervisor inspected representative drafts, identified systematic rule failures, and folded corrected rules back into the reviewer prompt. Examples included red-pixel signals caused by Pride flags or political-party banners, and anti-bullying advocacy mislabeled as harassment. At the final checkpoint, the supervisor adjudicated all \textsc{needs\_review} and \textsc{escalation} flags and re-examined high-impact category shifts, especially C9 upgrades and M7 thumbnail-mismatch flags. Final labels reflect supervised adjudication; independent dual annotation remains future work.

\subsection{Runtime Depth and Analysis}

To compare static and runtime evidence, we recorded the deepest level of exploration required to observe the primary safety-relevant signal. D0 indicates visibility from metadata; D1, visibility immediately after opening or starting; D2, appearance during passive runtime; D3, interaction, gameplay progression, or a failure state; and D4, source-code inspection, hidden asset inspection, or non-obvious trigger discovery. We summarize categories and risk levels, compare evidence channels and runtime depths, and analyze boundary disagreements such as horror versus violence, stylized combat versus gore, anti-awareness versus targeted harm, and political sensitivity versus high-risk content.

%% file: sec/evaluation.tex
\section{Findings}
\label{sec:findings}

The findings characterize evidence patterns in a targeted seed set that oversamples sensitive and borderline projects. The counts describe curation challenges in these pathways.

\subsection{Categories and Risk Were Related but Distinct}

The seed set contained horror, combat, gore, harassment, political themes, moderation-conflict signals, and possible high-risk content. Table~\ref{tab:category-risk-results} summarizes the observed category and risk distributions; categories are multi-label, so counts can exceed the number of projects. Category membership did not determine severity. Horror projects sometimes included violent threat framing or stylized blood-spatter assets. Combat projects ranged from cartoon mechanics to explicit gore. School-themed horror sometimes blurred fictional threat and targeted bullying. The C9 ``don't report'' cue often co-occurred with C2 horror or C6 harassment. These patterns support separating category from risk level.

\begin{table}[t]
\centering
\caption{Observed content categories and risk levels in the 500-project audit.}
\label{tab:category-risk-results}
\footnotesize
\setlength{\tabcolsep}{3pt}
\renewcommand{\arraystretch}{1.02}
\begin{tabularx}{\linewidth}{@{}X r@{}}
\toprule
Label & Projects \\
\midrule
\multicolumn{2}{@{}l}{\emph{Content categories (multi-label)}} \\
C1 Violence / Threat & 233 \\
C2 Horror / Survival & 252 \\
C3 Gore / Blood & 73 \\
C4 Sexual / NSFW & 2 \\
C5 Hate / Discrimination & 2 \\
C6 Harassment / Bullying & 27 \\
C7 Self-harm / Suicide & 6 \\
C8 Political / Societal & 99 \\
C9 Moderation Conflict / Report Avoidance & 25 \\
\midrule
\multicolumn{2}{@{}l}{\emph{Risk levels}} \\
0: No apparent risk & 3 \\
1: Sensitive but acceptable & 248 \\
2: Mildly age-inappropriate & 182 \\
3: Clearly inappropriate & 21 \\
4: High-risk or likely policy-violating & 46 \\
\bottomrule
\end{tabularx}
\end{table}

\subsection{Runtime Inspection Revealed Signals Missed by Static Review}

Exploration-depth and reveal-mechanism annotations provide the main evidence for the runtime-aware framing. Some signals were visible from metadata; many required execution, interaction, gameplay progression, failure states, or code and asset inspection. Table~\ref{tab:runtime-results} summarizes these observations.

\begin{table}[t]
\centering
\caption{Exploration depth and reveal mechanisms in the 500-project audit.}
\label{tab:runtime-results}
\footnotesize
\setlength{\tabcolsep}{3pt}
\renewcommand{\arraystretch}{1.02}
\begin{tabularx}{\linewidth}{@{}X r@{}}
\toprule
Label & Projects \\
\midrule
\multicolumn{2}{@{}l}{\emph{Exploration depth (per project, deepest)}} \\
D0: Visible from static metadata & 33 \\
D1: Visible immediately after opening or starting & 10 \\
D2: Appears during passive runtime & 70 \\
D3: Requires interaction, gameplay, or failure state & 189 \\
D4: Requires code or hidden asset inspection & 198 \\
\midrule
\multicolumn{2}{@{}l}{\emph{Reveal mechanisms (multi-label)}} \\
M0 Static / metadata-level & 407 \\
M1 Passive runtime & 342 \\
M2 Interaction-triggered & 328 \\
M3 Timed or delayed & 0 \\
M4 Broadcast / state-triggered & 190 \\
M5 Composite / hidden-asset reveal & 198 \\
M6 Audio-dominant & 4 \\
M7 Thumbnail / runtime mismatch & 28 \\
M8 Source-only or asset-only latent content & 0 \\
Unknown & 10 \\
\bottomrule
\end{tabularx}
\end{table}

Only 33 of 500 projects (7\%) were resolvable from static metadata at depth D0. The remaining 467 (93\%) required runtime exploration, and 387 (77\%) required interaction, gameplay progression, failure states, or hidden-asset/code inspection. The D3 and D4 cases show why project-page review is incomplete: a project can look benign under static inspection and reveal safety-relevant content through interaction, hidden assets, or event-triggered behavior. The 28 M7 cases further illustrate the gap: thumbnails or first frames diverged from later runtime behavior. Many runtime-revealed projects remain low risk, so runtime evidence supports calibrated curation instead of blanket exclusion.

Exploration depth also changes the feasibility of review. D1 and D2 evidence can often be surfaced by a short preview run. D3 evidence requires a reviewer or tool to try plausible learner actions, including clicks, movement, and failure states. D4 evidence requires asset or source inspection when the runtime path is unclear. These distinctions help educators and dataset builders report how much confidence their screening can support.
M3 and M8 remain in the scheme, although no project in this audit was adjudicated as primarily timed/delayed or source-only/asset-only; observed delays and latent assets co-occurred with stronger mechanisms.

\subsection{Disagreements Were Systematic}
\label{subsec:disagreements}

Annotator disagreements clustered around predictable boundaries. \textit{Horror versus violence}: monster-chase sequences were labeled C2 when fear and survival pressure dominated, and C1 when harm framing was explicit. \textit{Stylized combat versus gore}: red visual effects required runtime observation because intensity and audio context determined whether an effect read as stylized or graphic; rendered keyframes separated blood-related evidence from Pride flags, political-party banners, and Halloween title cards. \textit{Anti-awareness versus targeted harm}: anti-bullying and anti-racism advocacy projects were labeled C8 for societal sensitivity, while C5 and C6 were reserved for content that targeted an identifiable group or user. \textit{Contextual versus runtime evidence}: studio membership, remix-chain origin, and project-cluster context raised annotator attention, but category and risk assignment still required direct observation.

These disagreements identify where operational definitions matter most. A collapsed taxonomy would hide the distinctions that runtime-aware annotation is designed to capture.

%% file: sec/discussion.tex
\section{Discussion, Ethics, and Limitations}
\label{sec:discussion}

\subsection{Discussion}
Our 500-project audit frames sensitive-content screening in Scratch as runtime-aware educational curation for educators and computing education researchers preparing student-facing examples, browsing activities, remix materials, and datasets. In this targeted corpus, project-page screening left the primary safety-relevant signal unresolved for 467 projects (93\%); 387 projects (77\%) required interaction or deeper inspection. The practical implication is a screening-depth norm: curation records and dataset documentation should report whether review stopped at metadata, ran the project, tested interactions and failure states, or inspected hidden assets and code. The contribution is a vocabulary for runtime evidence and empirical support that static-only review is inadequate in this targeted corpus.

These evidence dimensions support several pathways. Teacher-selected examples need previewable runtime evidence; student-directed browsing needs guidance, safe exits, and shared norms; datasets need screening-depth provenance. For classroom use, a short workflow can record page cues, tested runtime paths, interactions, sounds, hidden costumes, and confidence. The goal is calibrated reuse: a mild horror project may fit an older elective but be unsuitable for open-ended browsing by younger students. When open exploration is part of the activity, runtime-aware curation can also teach learners to pause, exit, report, or ask for help after an unexpected audio or visual reveal.

For datasets and future tools, screening depth should be part of provenance: a dataset filtered only through titles and thumbnails cannot guarantee that runtime-revealed content was absent. A runtime preview assistant could run a project, exercise common inputs, surface hidden assets, flag thumbnail/runtime mismatches, and present evidence cards showing where a signal appeared. Teachers and researchers should remain in the decision loop because appropriateness depends on age group, assignment, and classroom norms; this aligns with youth-safety work emphasizing agency and collaborative guidance for minors~\cite{wisniewski2017middleground}. Follow-up studies should examine how teachers and students respond to runtime evidence during selection, browsing, and remixing.

\subsection{Ethics and Limitations}
Scratch is a youth-oriented platform, so we use a conservative approach to sensitive material. The analysis is restricted to publicly accessible projects. We avoid identifying or contacting creators, reporting usernames, and reproducing explicit imagery, personal information, or harmful text. Severe cases are described only in paraphrase and with minimal detail, and exploration halted upon encountering explicit sexual content, extreme gore, severe self-harm material, or real personal information.

Several limitations bound this work. The seed set deliberately oversamples sensitive and borderline projects; observed distributions are not prevalence estimates. Runtime exploration is time-bounded, so content gated behind extended gameplay or non-obvious interactions may remain undiscovered. Annotation requires judgment at category boundaries; the internal lead-researcher adjudication improves consistency but does not provide independent inter-rater reliability. This audit also leaves public-context evidence (E6), such as forum posts or moderation history, for future work. Because Scratch projects can be edited, removed, remixed, or recontextualized after collection, future studies should report sampling strategy, screening depth, and annotation confidence.

%% file: sec/conclusion.tex
\section{Conclusion}
\label{sec:conclusion}

This paper identifies runtime-revealed sensitive content as a computing education curation challenge for learner-facing Scratch reuse and dataset construction. Scratch projects are executable programs whose safety-relevant content can appear through interaction, timing, gameplay progression, failure states, hidden assets, or event triggers. We introduced a runtime-aware annotation scheme and used it to audit 500 public Scratch projects. The study shows why static project-page review is incomplete for a targeted corpus of executable youth media and motivates future work on classroom project selection, transparent dataset construction, and educator-facing screening tools.